\documentclass[a4paper,10pt,twoside]{cpc-hepnp}
\usepackage{multicol}
\usepackage{graphicx}
\usepackage{booktabs}
\usepackage{amssymb,bm,mathrsfs,bbm,amscd}
\usepackage[tbtags]{amsmath}
\usepackage{lastpage}

\begin{document}

\fancyhead[c]{\small Chinese Physics C~~~Vol. xx, No. x (202x) xxxxxx}
\fancyfoot[C]{\small 010201-\thepage}

\footnotetext[0]{Received xx xx 2021}

\title{Fast X-ray spectrum and image acquisition method for the XFEL facility\thanks{National Key Scientific Instrument and Equipment Development Project of China(Grant No. 11927805) and National Science Foundation for Young Scientists of China (Grant No. 12005134)}}

\author{%
      Shuo Zhang $^{1)}$\email{shuozhang@shanghaitech.edu.cn}
\quad Jing-Kai Xia $^{1}$%
\\
\quad Xu-Dong Ju $^{1}$%
}
\maketitle

\address{%
$^1$ {Center for Transformative Science, ShanghaiTech University, ShangHai, \quad 201210, China}
}

\begin{abstract}
X-ray free electron laser (XFEL) can provide X-ray light with about four order of magnitude higher flux than synchrotron radiation. Pulse light from XFEL interacts with the target and the resulting photons are collected by detectors. The strong intensity of XFEL will make multiple photons hit on one detector pixel and affect the photon energy measurement. Although increasing the distance between the target and detector could reduce the photons pile-up, it causes a waste of photons. So the traditional photon counting spectrum acquisition method is not advantageous in the XFEL case. To meet the requirements on both spectrum acquisition and imaging, we propose a new detection method in this paper.
\end{abstract}



\footnotetext[0]{\hspace*{-3mm}\raisebox{0.3ex}{$\scriptstyle\copyright$}2021
Chinese Physical Society and the Institute of High Energy Physics of the Chinese Academy of Sciences and the Institute of Modern Physics of the Chinese Academy of Sciences and IOP Publishing Ltd}%

\begin{multicols}{2}

\section{Introduction}

X-ray free electron laser provides X-ray light with ultra high intensity. Traditional photon counting detectors do not work well for the energy spectrum measurement in such high intensity case. The spectra acquisition with spatial resolution will take long time using traditional detection method.

In this paper, a spectra measurement method with spatial resolving ability using different X-ray filters is proposed.

\subsection{Classification of X-ray spectrometer}
Traditional X-ray spectrometer can be classified into energy-dispersive (ED) type and wavelength-dispersive (WD) type. The ED type has relatively worse energy resolution and longer integral time, but can used for imaging, while the resolution of WD type is high and imaging with WD type is difficult. The WD type is able to obtain the spectrum with single XFEL pulse, but the ED type is not.

We propose a new X-ray spectrum acquisition method different with the traditional two. In this method, X-ray filters and fast integrating detector are combined, to provide spectrum measurement and imaging ability with single pulse.

\subsection{Principle introduction}

\subsubsection{Mathematical calculation}
When X-ray photons pass a material M1 with thickness d1, absorption will happen with a probability P1. The transmission probability is $T1=1-P1$, with P1 relates to d1, the elements composition Z1 of the material, and the energy E of the X-ray photons. For given d1 and Z1, there is a relationship $T1_{(E)}$ between T1 and E. For different materials, it can be written as $Tn_{(E)}$. If N*N-1 different filters are placed in front of N*N detector pixels, with the first pixel not filtered, different incident energy spectra on each pixel will result in N*N different signal amplitudes.

If the original incident spectrum is $S_{(E)}$, the integrated charge on the first and number n pixel are:
\begin{equation}\label{eq1}
  Q_0=\int_{E_{min}}^{E_{max}}\frac{S_{(E)}*1*E}{E_{0}}*\delta{E}
\end{equation}

\begin{equation}\label{eq2}
  Q_n=\int_{E_{min}}^{E_{max}}\frac{S_{(E)}*Tn_{(E)}*E}{E_{0}}*\delta{E}
\end{equation}

There will be N*N equations. The probability density in N*N energy bins could be calculated via a N*N Q value matrix. But in practical, the calculation of basis vectors is difficult and the statistical fluctuations will affect the results.

Since most natural materials have smooth $T_{(E)}$ curves instead of band-pass property, except for significant change near the absorption edge, the calculation of basis vectors for the N*N matrix is complicated. The statistical fluctuation effect in the photon absorption and detection process, and also in the charged electron-hole generation, will distort the measured spectra.

\subsubsection{Basis vectors definition}
A energy spectrum is the statistical data on different energy bins in a certain range, so the number of bins directly affect the precision of the spectrum. In the proposed detection method, the number of energy bins can be increased by increasing the number of N, to get better spectrum precision. However, there are some limitations. The first is the finite number of natural elements in filters, which limits the energy selection even if different thickness is used. The second is the spatial resolution will be worse with large N. The third is the interested energy line number in application is small usually, only two to three sometimes. So too many energy bins may not give more information, but make the matrix calculation difficult. The last limitation is the count in each bin reduces while the bins number increases, and give larger statistical fluctuation. So too large bins number maybe not necessary, and we use N=3 as an example in the following discussion.

\begin{center}
\includegraphics[width=8cm]{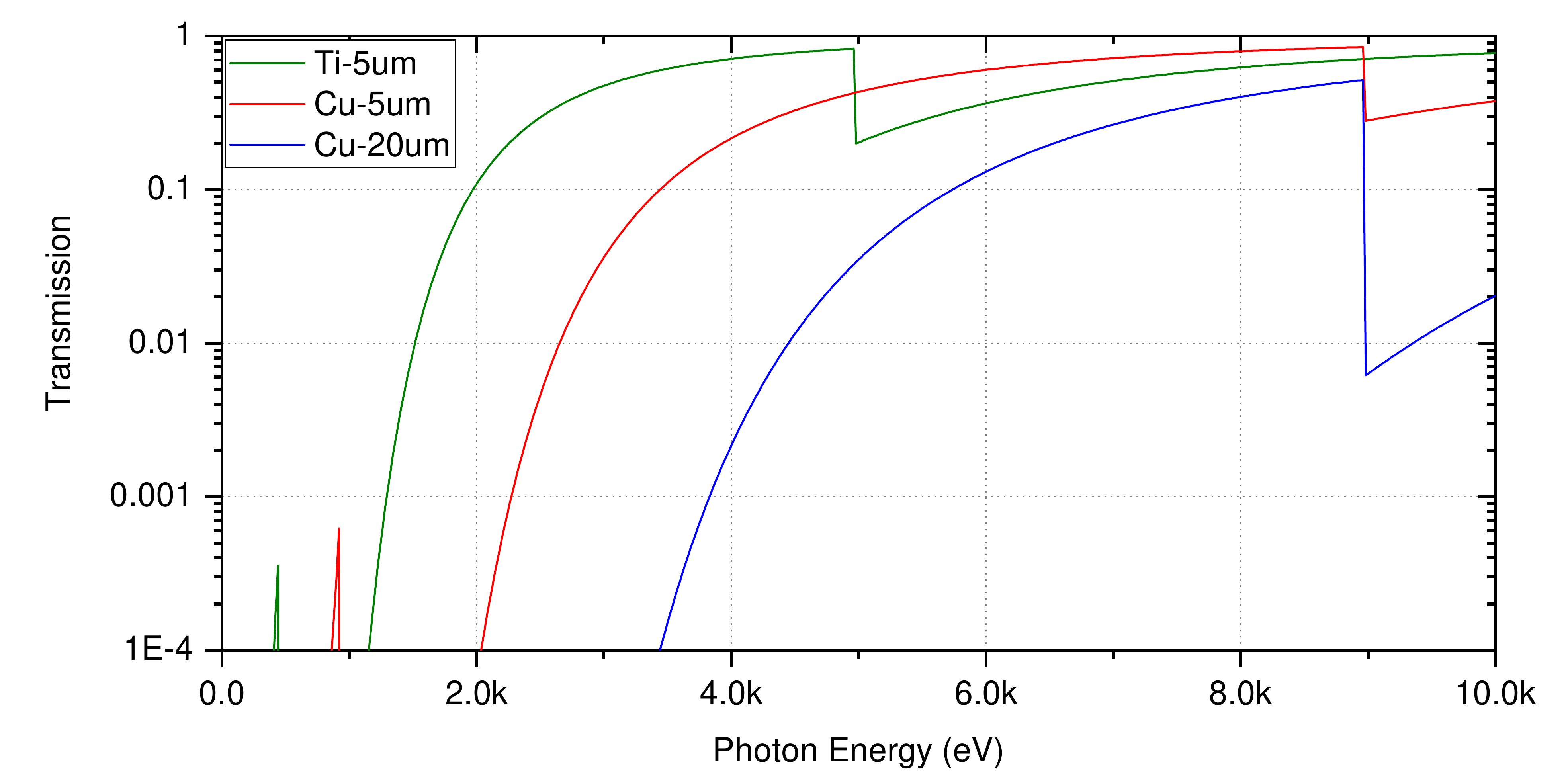}
\figcaption{\label{fig1}   Transmission curves for filters with different elements composition and thickness.}
\end{center}

To simplify the calculation, here we assume we can find nine basis vectors which are the nine parts of the energy space. The transmission matrix of each part is a diagonal matrix of $E_i$, $Tn_{(E_i)}=\delta_{n,i}$. This means there are eight filters used for band-pass without gap and overlap. The ninth pixel is not filtered, and can make the ninth basis vector after substraction with the other eight vectors. In \ref{fig2}, it can be seen that several interested energy lines locate in a same filter bin. This problem can be solved by changing the material of the filter or increasing the number of the basis vectors. In reality, a filter curve is the combination of a number of basis vectors.

\begin{center}
\includegraphics[width=8cm]{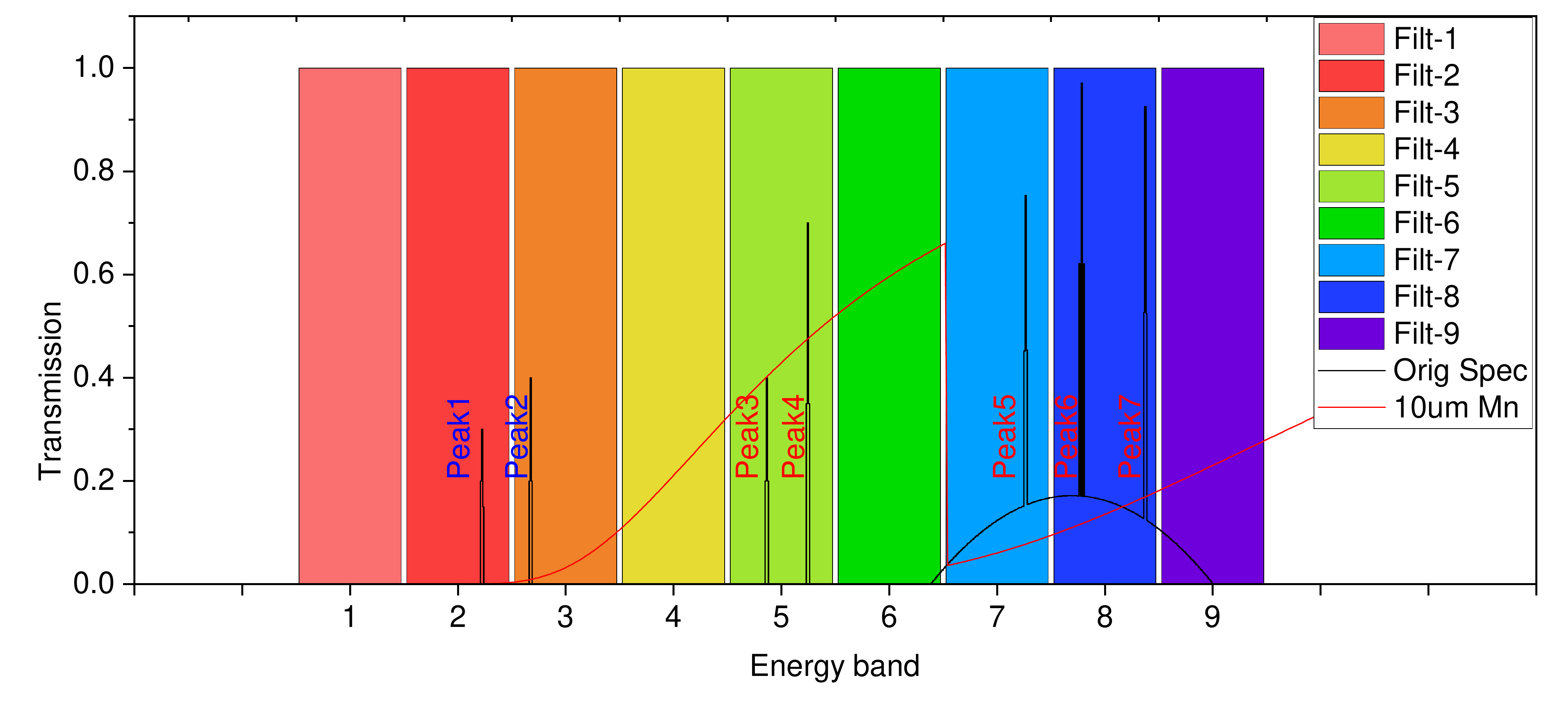}
\figcaption{\label{fig2}   Comparison of the transmission curves by 10$\mu$m Mn and nine ideal basis vectors.}
\end{center}

The original spectrum is $N_0*s(E_i)$, where $s(E_i)$ is the density distribution after normalization and $N_0$ is the number of total photons. Using $N_0*s(E_i)$ in \ref{eq2} will give:
\begin{equation}\label{eq3}
  Q_n=N_0*\sum_{E_{1}}^{E_{9}}\frac{s_{(E_i)}*\delta_{n,i}*E_i}{E_{0}}*\delta{E}
\end{equation}
or
\begin{equation}\label{eq4}
  Q_n=(N_0*\overline{s_{(E_i)}})*\frac{E_i}{E_{0}}
\end{equation}

\subsubsection{Statistical fluctuation effect on the energy resolution}

The uncertainty of \ref{eq4} is
\begin{equation}\label{eq5}
  \Delta{Q_n}=\delta(N_0*\overline{s_{(E_i)}})*\frac{E_i}{E_{0}}+(N_0*\overline{s_{(E_i)}})*\delta\frac{E_i}{E_{0}}
\end{equation}
\begin{equation}\label{eq6}
  \Delta{Q_n}/Q_n=\delta(N_0*\overline{s_{(E_i)}})/(N_0*\overline{s_{(E_i)}})+\delta\frac{E_i}{E_{0}}/\frac{E_i}{E_{0}}
\end{equation}

So in the simplified case, the uncertainty in the bin $E_i$ is decided by both the number of incident photons and the electron-hole pair generated by a single photon. If the incident light flux is large enough, the uncertainty will be mainly limited by the detector resolution. For the semiconductor detector, the resolution is limited by the Fano factor, which is close to 2$\%$ around 6keV. In this condition, increasing the incident photons number on a single pixel will not obtain remarkable improvement on accuracy when it is larger than 10000. This is because for the integrating mode in the method, the photon counting results in a total charge value Q. If the number of incident photons is small, its fluctuation $\delta{N}$ will cause larger uncertainty on the signal amplitude $\delta{N}*\frac{E_i}{E_{0}}$ than that from the fluctuation of the number of generated electron-hole pairs, $N*\delta\frac{E_i}{E_{0}}$. When the incident flux get large enough, the term $\delta\frac{E_i}{E_{0}}/\frac{E_i}{E_{0}}$ will contribute to the main uncertainty.

The second term of \ref{eq6} set strong limitation in high background situation. If the intensity of the interested line is only one percent of the background, it will be submerged in the fluctuation of background.

To effectively make use of photons and reduce the statistical uncertainty, nine pixels may be enough in low repetition rate or synchrotron radiation case. The center pixel is not filtered and read out by a separate ADC with eight times or higher rate compared to other pixels for averaging. The other eight pixels require longer photon collection time as the filtering effect, so they can share a ADC. The nine pixels are called a pixel group in the following discussion.

\ref{eq6} partially explains the necessity of limitations required in some imaging measurements. For example, the incident light should be monochromatic, to make the signal amplitude be proportional to the photon numbers. In our case, the light could not be monochromatic. By reducing the area of the pixel group and increasing the group number, the number of incident photons can be estimated from amplitudes averaging among pixel groups. This is because the pixel in the neighbouring group can take repeated measurement, which improves the accuracy.

Increasing the number of filters can also reduce the effect of statistical fluctuation. If increase the pixel number from 9 to 25, or even 49, the additional pixels will share photons, which is similar to multiple measurements. The reason of measure the center pixel eight times is also for reducing uncertainty by fluctuation.

In addition, sensors like STJ or microcalorimeters which have three to four lower mean ionization energy than the semiconductor maybe useful to reduce the limitation of the second term in \ref{eq6}, obtaining lower fluctuation by collecting more photons. The difficulty is STJ and microcalorimeters work in low temperature like 0.3K or lower, which means intensive incident X-ray will cause heavy heat load to the refrigerator by heating the detector.  So this method need further detail consideration.

\subsubsection{Practical case and simulation}
Many assumptions are set in previous discussions, especially on the basis vectors. However, it is not possible to find out eight ideal band-pass filters without gap and overlap in reality. Since the final energy resolution relates to the incident light flux, the composition and thickness of the filters, and the structure of the original spectrum, the expected result should be calculated combining these parameters. \ref{fig3} show the calculated relative Q values in each pixel. It shows that the spectrum variation around the absorption edge of Co can be found via the change of Q values.
\begin{center}
\includegraphics[width=8cm]{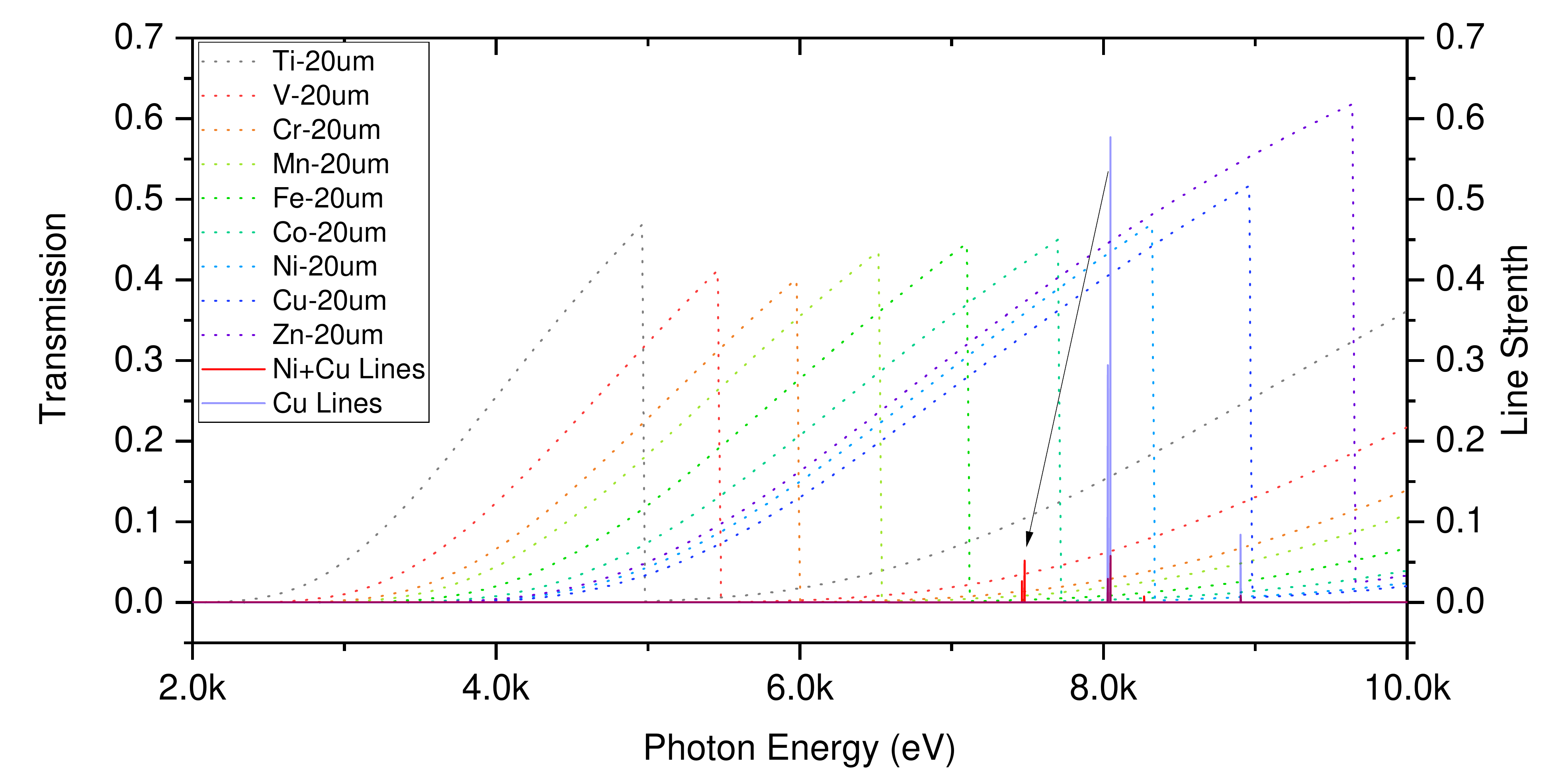}
\figcaption{\label{fig3}   The transmission curves of 20$\mu$m thick filters made by different elements, the characteristic X-ray emission lines of Cu and their change after interaction with a Ni sample.}
\end{center}
\begin{center}
\includegraphics[width=8cm]{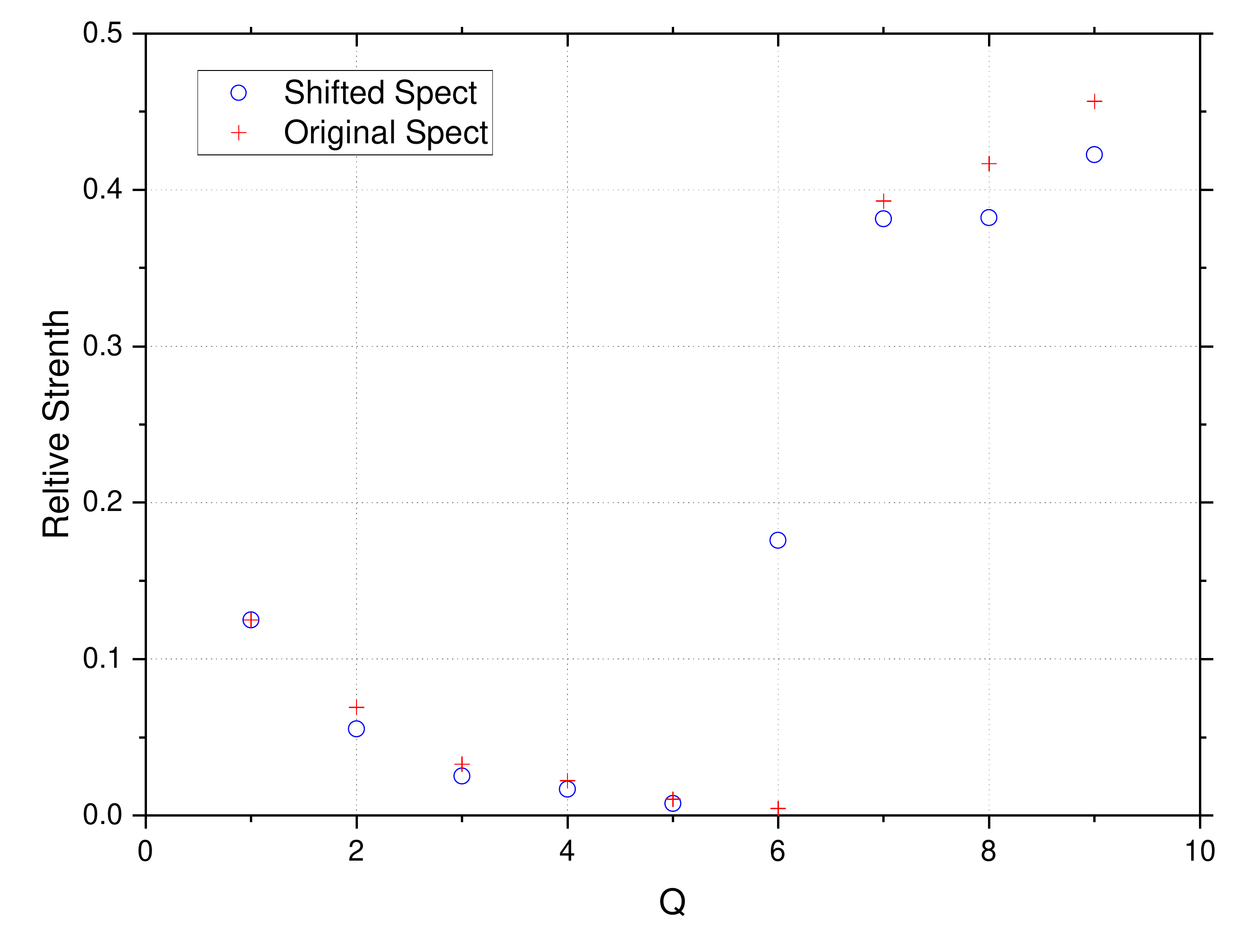}
\figcaption{\label{fig4}   The relative Q value in each pixel with the characteristic X-ray emission lines in \ref{fig3} passing different filters. The first pixel is not filtered, but its Q value is averaged by 8.}
\end{center}

Although the practical transmission curves are not ideal basis vectors, they can be demerged to several ideal basis vectors. So the practical measurement increases the complication of calculation, but still limited by \ref{eq6} in uncertainty.

\section{Conclusion}
In principle, the proposed method can not give accurate calculation on the incident energy spectrum via the measured Q value. Possible solutions to reduce the statistical effect include increasing the pixel group number, increasing the pixel number in a group, using other sensors like microcalorimeters and STJ which have low mean ionization energy.

The proposed method is useful for energy measurement and imaging with single pulse in some specific cases since the result is sensitive to the variation of incident spectrum. Although the accurate spectrum can not be identified, the profile of the incident photon energy can be obtained, which is helpful in some applications. In addition, this method is efficient in identifying multi-photons.
\section{acknowledgment.}
We acknowledge the support of National Key Scientific Instrument and Equipment Development Project of China(Grant No. 11927805) and National Science Foundation for Young Scientists of China (Grant No. 12005134).

\vspace{-1mm}
\centerline{\rule{80mm}{0.1pt}}

\end{multicols}{2}
\clearpage
\end{document}